\documentclass[epj]{svjour}

\usepackage{amssymb,amsmath,bm}
\usepackage{graphicx}
\usepackage[colorlinks]{hyperref}
\usepackage{graphicx}
\usepackage{subfigure}

\usepackage{xcolor}
\usepackage[normalem]{ulem}  

\usepackage[normalem]{ulem}

\providecommand{\apj}{Astrophys.\ J. }

\providecommand{\apjs}{Astrophys.\ J.\ Suppl. }
\providecommand{\aap}{Astron.Astrophys. }

\providecommand{\prc}{Phys.\ Rev.\ C }
\providecommand{\prd}{Phys.\ Rev.\ D }
\providecommand{\prl}{Phys.\ Rev.\ Lett.\ }
\providecommand{\plb}{Phys.\ Lett.\ B }

\providecommand{\npa}{Nucl..\ Phys.\ A }

\begin{document}
\bibliographystyle{ieeetr}
\title{Consequences of simultaneous chiral symmetry breaking and deconfinement for the isospin symmetric phase diagram\thanks{Contribution to the Topical Issue ÒExploring strongly interacting matter at high densities - NICA White PaperÓ edited by David Blaschke et al.}}
\titlerunning{Simultaneous chiral symmetry breaking and deconfinement}
\author{Tobias Fischer\inst{1,}\thanks{\emph{email:} fischer@ift.uni.wroc.pl}, Thomas Kl{\"a}hn\inst{1} and Matthias Hempel\inst{2}}
\institute{
Institute of Theoretical Physics, University of Wroclaw, Pl. M. Borna 9, 50-204 Wroclaw, Poland,
\and
Department of Physics, University of Basel, Klingelbergstrasse 82, 4056 Basel, Switzerland
}
\date{Received: 18 March 2016 / Revised: 16 June 2016 \\
Published online: 16 August 2016 \\
\copyright\,The Author(s) 2016. This article is published with open access at Springerlink.com Communicated by D. Blaschke
}
\abstract{
The thermodynamic bag model (tdBag) has been applied widely to model quark matter properties in both heavy-ion and astrophysics communities. Several fundamental physics aspects are missing in tdBag, e.g., dynamical chiral symmetry breaking (D$\chi$SB) and repulsions due to the vector interaction are both included explicitly in the novel vBag quark matter model of Kl{\"a}hn and Fischer~(2015)~(Astrophys.\ J.\ {\bf 810},134 (2015)). An important feature of vBag is the simultaneous D$\chi$SB and deconfinement, where the latter links vBag to a given hadronic model for the construction of the phase transition. In this article we discuss the extension to finite temperatures and the resulting phase diagram for the isospin symmetric medium.
%
     } 
%
\maketitle

\section{Introduction}
\label{intro}

The theory of strong interactions, i.e. Quantum Chromodynamics (QCD), considers hadrons and mesons as color neutral compound objects of quarks and gluons. Besides confinement a second key feature of QCD is dynamical chiral symmetry breaking (D$\chi$SB) and its restoration at large densities and high temperatures. Currently, lattice-QCD is the only ab-initio approach to solve QCD numerically~(cf. Refs.~\cite{Fodor:2004nz,Aoki:2006we} and references therein) applicable in the vicinity of vanishing chemical potentials. A smooth cross-over phase transition is predicted at $154\pm9$~MeV (cf. Refs.~\cite{Katz:2012JHEP,Laermann:2012PRD,Laermann:2012PRL,Katz:2014PhLB} and references therein). It is in qualitative agreement with heavy-ion collision experiments \cite{Redlich:2015}. In the latter at moderate and low collision energies one encounters finite chemical potentials (or equivalently high densities around normal nuclear density) and high temperatures which are currently inaccessible for lattice QCD.

Medium properties of quark matter have long been studied (cf. Refs. \cite{Bender:1997jf,Roberts:2000aa,Buballa:2003qv,Alford:2006vz,Pagliara:2007ph,McLerran:2007qj,Sagert:2008ka,Pagliara:2009dg,Blaschke:2009,Klaehn:2010,Klahn:2011fb,Chen:2011my,Fischer:2011,Weissenborn:2012,Bonnano:2012,Blaschke:2013zaa,Kurkela:2014vha,Benic:2014jia,Schaffner:2014} and references therein). The two most commonly used effective quark matter models are the thermodynamic bag model (tdBag) of Ref.~\cite{Farhi:1984qu} and models based on the Nambu-Jona-Lasinio (NJL) approach~\cite{Nambu:1961tp,Klevansky:1992qe,Buballa:2003qv}. Recently, we illustrated in Ref.~\cite{Klaehn:2015} that both approaches can be understood as limiting solutions of QCD's in-medium Dyson-Schwinger gap equations \cite{Bashir:2012fs,Cloet:2013jya,Chang:2011vu,Roberts:2012sv,Roberts:2000aa,Chen:2008zr,Chen:2011my,Chen:2015mda,Klahn:2009mb} within a particular set of approximations. Note that  perturbative QCD is only valid in the limit of asymptotic freedom, i.e. where quarks are no longer strongly coupled~\cite{Kurkela:2014vha}.

Currently, no consistent approach exists to describe medium properties of hadrons and mesons at the level of quarks and gluons at high density. Hence, the deconfinement phase transition is usually constructed from a given hadronic EoS with hadrons and mesons as the fundamental degrees of freedom and an independently computed quark matter EoS. Constructions based on Maxwell's and Gibbs conditions by definition result in a 1st-order phase transition. 

We introduced the novel quark matter EoS vBag~\cite{Klaehn:2015} in order to account explicitly for both D$\chi$SB and repulsive vector interactions. The latter property is essential for a stiffening of the EoS towards high densities, known already from NJL. Moreover, vBag mimics (de)confinement via a phenomenological correction to the EoS. It is determined by the hadronic EoS chosen for the construction of the phase transition and it ensures that chiral and deconfinement phase transitions coincide. That this might be the case has been suggested by different Dyson-Schwinger studies (cf. Ref.~\cite{Qin:2010nq,FischerC:2014}). We point out, that connecting a nuclear and quark EoS in terms of a Maxwell construction always assumes simultaneous DCSB and confinement. The special feature of our approach is to assume that the critical chemical potential for DCSB is obtained from the quark matter model and thus defines the onset of confinement as well.

In this article we apply vBag as a 2-flavor model for isospin symmetric matter and discuss the corresponding phase diagram with potential applications of heavy-ion collisions in the energy range of NICA in Dubna (Russia) and FAIR at the GSI in Darmstadt (Germany). It corresponds to temperatures on the order of less than 100 MeV at densities around normal nuclear matter density. There, a 1st-order phase transition is expected, being in agreement with vBag by construction. A consequence of our model prescription is that the deconfinement bag constant $B_{\rm dc}$ becomes medium dependent. It is closely related to the hadronic EoS. For illustration we select a nuclear EoS from the catalogue of Ref.~\cite{Hempel:2009mc}. It treats the nuclear medium within a quasi-particle picture of the relativistic mean-field framework for conditions in excess of normal nuclear matter density and high temperatures. Specifically, we choose the parametrization DD2 from Ref.~\cite{Typel:2009sy}.

The manuscript is organized as follows. In sec.~\ref{vbag} we briefly review vBag and discuss phase transition as well as phase diagram in sec.~\ref{phase}. The manuscript closes with a summary in sec.~\ref{summary}.

\section{vBag}
\label{vbag}

The zero-temperature and single-flavor ($f$) quark matter model vBag is defined by the following set of equations for pressure ($P$), energy density ($\varepsilon$) and particle density ($n$),
\begin{eqnarray}
\mu_f &=& \mu_f^*+K_v n_{{\rm FG},f}(\mu_f^*)~,
\label{eq:vbagmu} \\
P_f(\mu_f) &=& P_{{\rm FG},f}^{kin}(\mu_f^*) + \frac{K_v}{2}  n_{{\rm FG},f}^2(\mu_f^*) - B_{\chi,f}~,
\label{eq:vbagP} \\
\varepsilon_f(\mu_f) &=& \varepsilon_{{\rm FG},f}^{kin}(\mu_f^*) + \frac{K_v}{2}  n_{{\rm FG},f}(\mu_f^*)^2 + B_{\chi,f}~,
\label{eq:vbage} \\
n_f(\mu_f) &=& n_{{\rm FG},f}(\mu_f^*)~,
\label{eq:vbagn}
\end{eqnarray}
in terms of Fermi-gas (FG) expressions. Moreover, the constant $K_v$ in Eqs.~\eqref{eq:vbagmu}--\eqref{eq:vbage} relates directly to the coupling strength of repulsive vector-current interactions as one would define them within NJL. This aspect extends beyond tdBag via the appearance of the effective chemical potential $\mu_f^*$ which enters all Fermi-gas expressions. The actual chemical potential $\mu_f$ -- in the sense of a thermodynamic variable -- is determined post-priori. This is common for quasi-particle models where interactions dynamically alter particle in-medium properties, e.g., mass and chemical potential but not the formal structure of the Fermi-gas integrals.

Note that vBag is valid only in the chirally restored phase assuming that bare quark masses sufficiently well approximate the mass gap solutions to describe the thermodynamical behavior of chirally restored quark matter. Further, it is useful to redefine vBag in terms of the baryon chemical potential $\mu_{\rm B} = \mu_u+2\mu_d$ where $\partial P/\partial \mu_{\rm B} = n_{\rm B}$ relates the pressure derivative with respect to the baryonic chemical potential to the baryon density $n_B$. We denote our critical chemical potential for D$\chi$SB  as $\mu_{\rm B,\chi}$, which is determined via the condition $\sum_f P_f=0$. It accounts for the breaking of chiral symmetry in confined matter and subsequently the restoration of chiral symmetry induces pressure to the system. For $\mu_{\rm B}<\mu_{\rm B, \chi}$ we assume that quarks are confined in hadrons and mesons which are not accessible for vBag. The chiral phase transition and the corresponding critical chemical potential $\mu_{\rm B,\chi}$ is defined by the value of the chiral bag constant $B_{\chi,f}$ in \eqref{eq:vbagP} and \eqref{eq:vbage}, demanding that at $\mu_{\rm B} = \mu_{\rm B,\chi,}$ the total pressure turns positive. In vBag $B_{\chi,f}$ is a parameter which can be computed from a microscopic approach as the difference between the vacuum pressure of the chirally broken and the chirally restored phases.

With the trivial extension of Eqs.~\eqref{eq:vbagmu}--\eqref{eq:vbagn} to finite temperatures -- note that temperature enters only the Fermi-gas expressions -- the entropy can be defined as follows,
\begin{eqnarray}
s_f(T,\mu_f) &=& \left.\frac{\partial P_f(T,\mu_f)}{\partial T}\right| _{\mu_f}=s_{{\rm FG},f}(T,\mu_f^*)~.
\end{eqnarray}
Here we assume $B_{\chi,f}$ to be medium independent.

In this work we assume that at least at sufficiently low temperatures the assumption of a first order transition applies. The condition for the transition from hadronic ($\mathcal{H}$) to quark matter ($\mathcal{Q}$) for the total pressure of each phase reads $P^\mathcal{H}(\mu_{\rm B,\rm dc})=P^\mathcal{Q}(\mu_{\rm B,\rm dc})$ at given $T$. Consequently one observes that chiral and deconfinement transitions -- the latter is defined via the deconfinement chemical potential $\mu_{\rm B,\rm dc}$ -- are located at different chemical potentials, i.e. $\mu_{\rm B,\chi}<\mu_{\rm B,\rm dc}$. In order to cure this inconsistency vBag accounts for simultaneous chiral symmetry breaking and confinement at $\mu_{\rm B,\chi}=\mu_{\rm B, \rm dc}$ by {\bf adding} the deconfinement bag constant $B_{\rm dc}$ to the total quark pressure as follows,
\begin{eqnarray}
&& P^Q=\sum_f P_f(T,\mu_f) + B_{\rm dc}~.
\end{eqnarray}
In order to match quark and hadron pressures one finds $B_{\rm dc}=P^\mathcal{H}(\mu_{\rm B,\chi})$ at given $T$. Hence $B_{\rm dc}$ depends on the nuclear EoS. This procedure effectively lowers the onset of deconfinement and ensures that quark matter is favored above hadronic matter for $\mu_{\rm B}>\mu_{\rm B,\chi}$. Alternative approaches are discussed in the literature~\cite{McLerran:2007qj,McLerran:2009,Blaschke:2010}, e.g., in terms of a quarkyonic phase where $\mu_{\rm B,\chi}\neq\mu_{\rm B, \rm dc}$.

According to our prescription for the phase transition it will vary with temperature: $B_\text{dc}\to B_\text{dc}(T)$, while we keep it constant with respect to $\mu_{\rm B}$. This medium dependence of $ B_{\rm dc}$ results in an additional contribution to the entropy density, $\partial B_{\rm dc}/\partial T=:s_{\rm dc}$, such that:
\begin{eqnarray}
&& s^Q =  \sum_{f=(u,d)} s_f + s_{\rm dc}~.
\end{eqnarray}
While the procedure does not induce a deconfinement baryon density, the energy density obtains additional contributions from $s_{\rm dc}$ in order to ensure thermodynamic consistency,
\begin{eqnarray}
\varepsilon^Q &=& \sum_{f=(u,d)} \varepsilon_f - B_\text{dc}(T) + T s_{\rm dc}~.
\end{eqnarray}
The functional dependence of $s_{\rm dc}$ is derived in Ref.~\cite{Klaehn:2016}. Note that at $T=0$ holds $s^\mathcal{H}=0$ and $\sum_f s_f=0$, from which follows $s_{\rm dc}(T=0)=0$. It is evident that the definition of $B_{\rm dc}(T):=P^\mathcal{H}(T,\mu_{\rm B,\chi})$ entangles vBag with the hadron EoS. 

Note that in analogy to the temperature also arbitrary isospin asymmetry and the associated charge chemical potential, $\mu_C=\mu_u-\mu_d$, induces additional corrections of $B_{\rm dc}$ to the EoS. This dependence and the corresponding deconfinement terms have been derived and discussed in details in Ref.~\cite{Klaehn:2016}. However, here we focus on symmetric matter where correction terms that relate to $\mu_C$ vanish, $B_{\rm dc}(T,\mu_C=0)=B_{\rm dc}(T)$ and $\partial B_{\rm dc}/\partial \mu_C=0$. 

\section{Phase Diagram}
\label{phase}

\begin{figure}[htp!]
\centering
\includegraphics[width=0.5\textwidth]{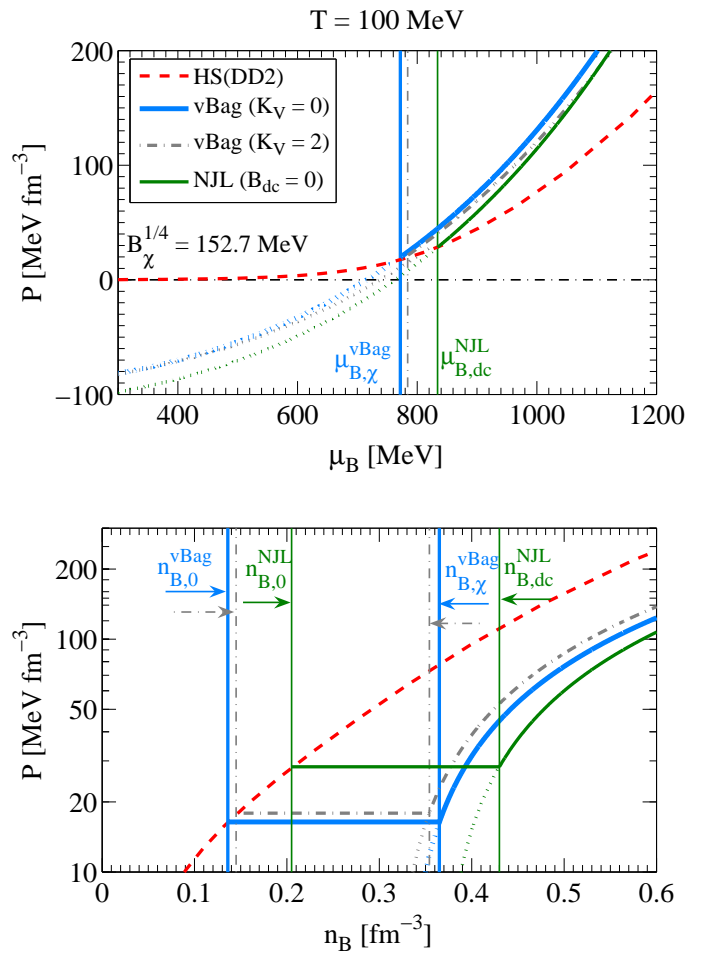}
\caption{(color online) Construction of the phase transition for a selected chiral bag constant $B_\chi$ comparing vBag  -- without ($K_V=0$) and with ($K_V=2$) vector interactions -- and the 'standard' NJL approach ($B_{\rm dc}=0, K_V=0$). {\em Top panel:} Dependence on baryochemical potential. {\em Bottom panel:} Dependence of the baryon density (see text for definitions).}
\label{fig:eos}
\end{figure}

\begin{figure*}[htp!]
\centering
\subfigure[~vBag]{
\includegraphics[width=0.485\textwidth]{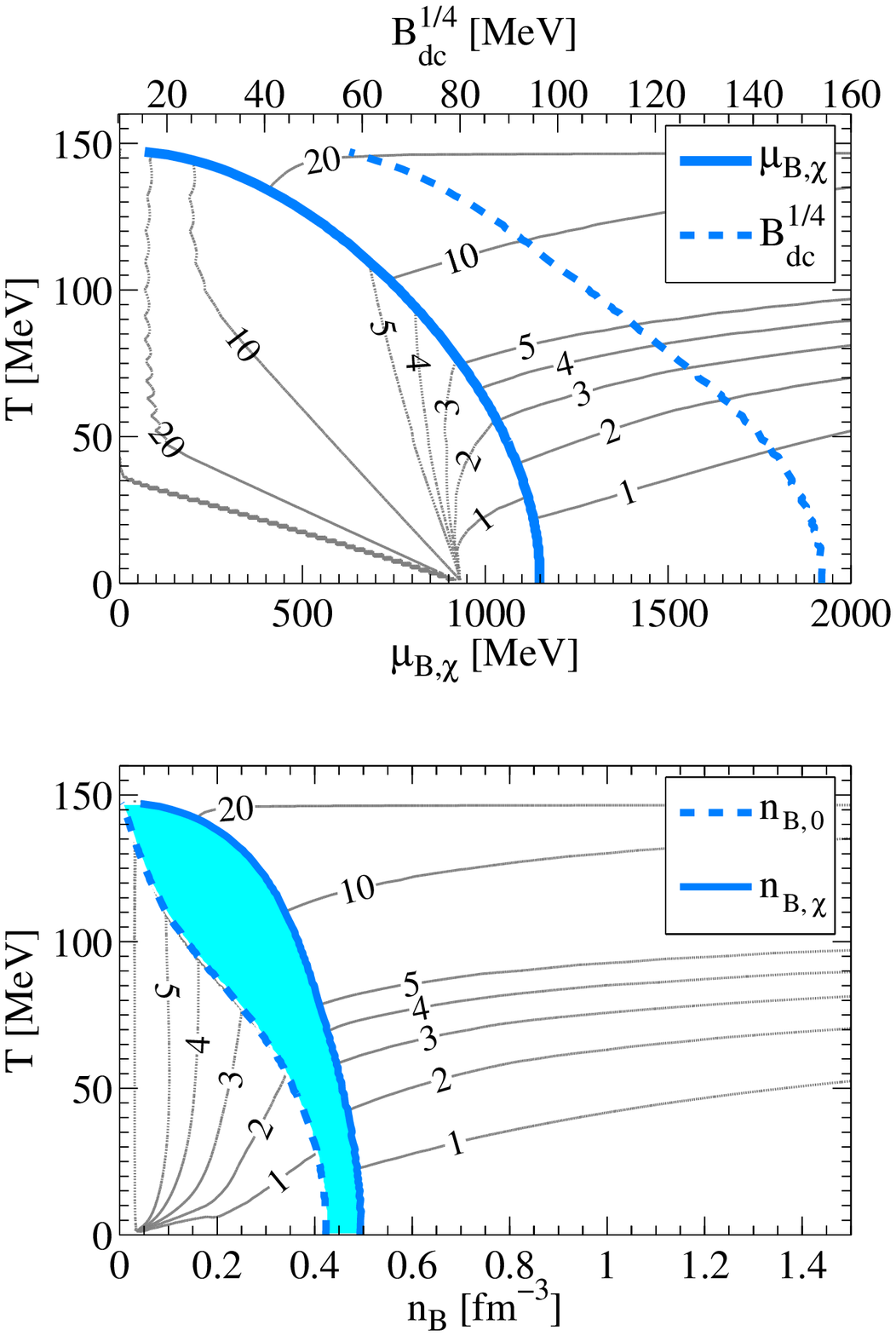}
\label{fig:phase_vbag}}
\hfill
\subfigure[~'standard' NJL approach]{
\includegraphics[width=0.485\textwidth]{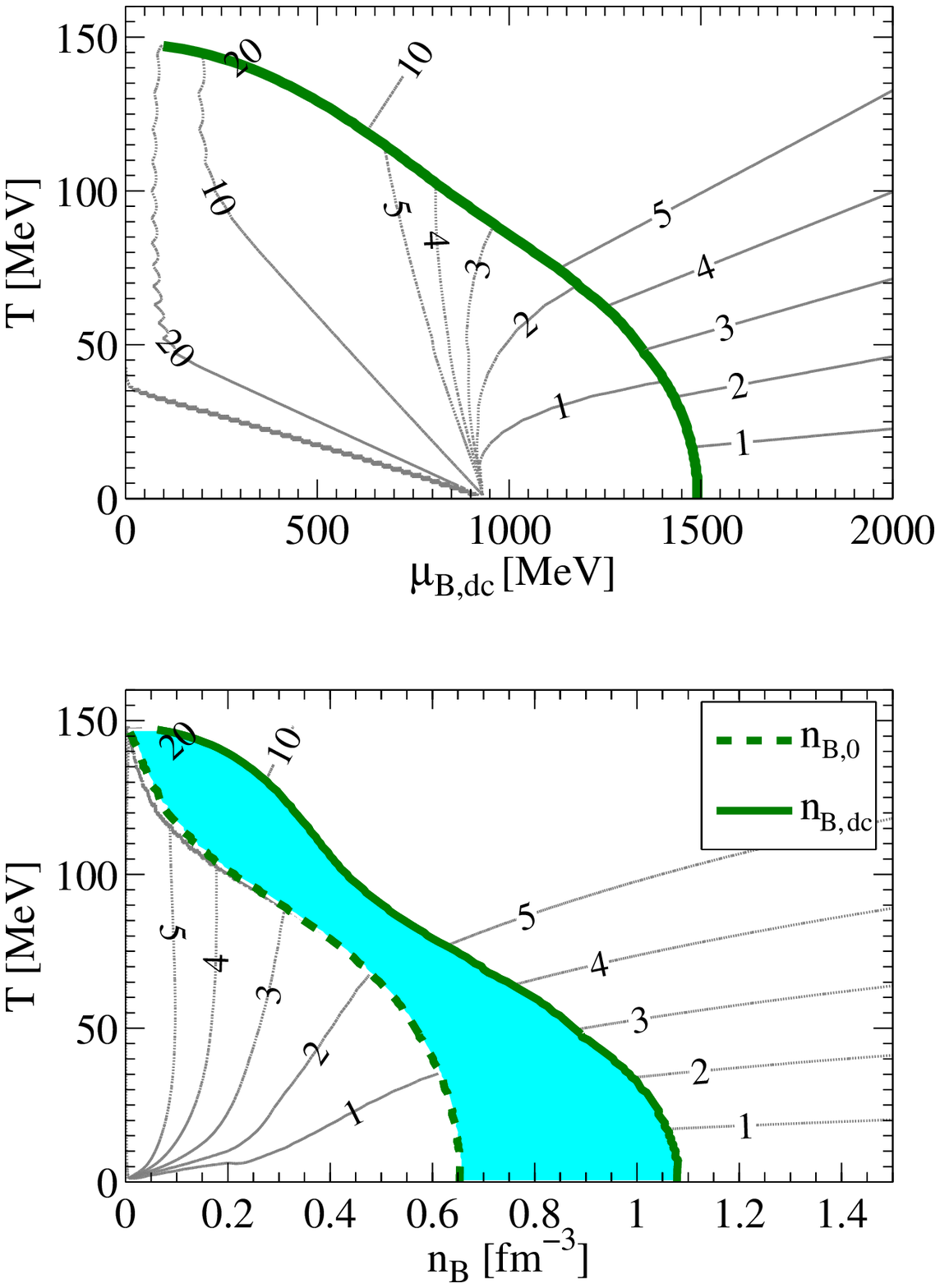}
\label{fig:phase_njl}}
\caption{(color online) Phase diagram for a selected chiral bag constant $B_\chi^{1/4}=152.7$~MeV, comparing our and the `standard' NJL approach ($B_{\rm dc}=0$). {\em Top panel:} Dependence on baryochemical potential. {\em Bottom panel:} Dependence {\bf on} the baryon density (see text for definitions). Solid gray lines mark lines of constant entropy per baryon in units of $[$k$_{\rm B}]$.}
\label{fig:phase}
\end{figure*}

In the following we discuss symmetric 2-flavor matter for which Figs.~\ref{fig:eos} and \ref{fig:phase} illustrate the phase transition construction and phase diagram assuming a chiral bag constant of $B_\chi^{1/4}=152.7$~MeV. The latter is in agreement with hadron physics and resembles the vacuum pion mass and form factors~\cite{Grigorian:1999}. It results in a chiral phase transition at the critical baryochemical potential of $\mu_{\rm B,\chi}=772$~MeV here for $T=100$~MeV characterized by $P^\mathcal{Q}>0$. The further pressure increase due to $B_{dc}$ ensures the deconfinement phase transition at $\mu_{B,\chi}$ as illustrated in the upper panel of Fig.~\ref{fig:eos}. As hadronic model we select the well constrained nuclear EoS of Ref.~\cite{Hempel:2009mc} (HS) which is based on the relativistic mean field description of the nuclear medium using the parametrization DD2~\cite{Typel:2009sy} henceforth denoted as HS(DD2). In the following we compare vBag with the commonly used approach for the construction of a 1st-order phase transition, i.e. $B_{\rm dc}=0$, which we denote as 'standard' NJL approach.

The top panel in Fig.~\ref{fig:eos} shows the  construction of the phase transition in the thermodynamic variables pressure and baryon chemical potential $P(\mu_{\rm B})$, for a selected temperature of 100~MeV. Vertical lines mark the position of the chiral phase transition for vBag (blue lines) and for the deconfinement transition for the 'standard' NJL approach (green lines). The difference between the latter approach and vBag becomes evident in the location of the phase transition at $\mu_{B,\rm dc}^{\rm NJL}=834$~MeV for $T=100$~MeV which significantly exceeds $\mu_{\rm B,\chi}^{\rm vBag}$. In particular at low temperatures both approaches result in increasing differences between $\mu_{\rm B,\chi}$ and $\mu_{\rm B,\rm dc}$. The corresponding phase transition in terms of the baryon density $n_B$ is shown in the bottom panel of Fig.~\ref{fig:eos}. The density where pure quark matter is obtained is marked by a vertical line $n_{\rm B,\chi}$ for vBag, corresponding to $\mu_{\rm B,\chi}$, and $n_{\rm B,\rm dc}$ for the 'standard' NJL approach, corresponding to $\mu_{\rm B,\rm dc}$. The phase transition is characterized by an extended density region in which hadrons and quarks co-exist between the onset density $n_{\rm B,0}$ and $n_{\rm B,\chi}^{\rm vBag}$ as well $n_{\rm B, \rm dc}^{\rm NJL}$ for the 'standard' NJL approach, at a constant pressure. In terms of density the 'standard' approach results in a later onset of the quark phase transition and of the pure quark phase compared to vBag.

From Fig.~\ref{fig:eos} it becomes also evident that the inclusion of vector interactions (here we select $K_V=2$ in units of $[10^{-10}$ MeV$^{-2}]$) has only little impact on the position of $\mu_{\rm B,\chi}$ (gray dash-dotted lines in Fig.~\ref{fig:eos}). It stiffens the EoS towards larger $\mu_{\rm B}$. The inclusion of vector interactions shrinks the region between $n_{\rm B,0}$ and $n_{\rm B,\chi}$ only mildly.

The size of the density coexistence region -- between $n_{\rm B,0}$ and $n_{\rm B,\chi}$/$n_{\rm B,\rm dc}$ -- depends on temperature. It is illustrated as light blue shaded region in the phase diagram at the bottom panels of Fig.~\ref{fig:phase}, comparing again vBag (Fig.~\ref{fig:phase_vbag}) and the 'standard' phase transition approach denoted as NJL (Fig.~\ref{fig:phase_njl}). In addition to a generally more extended phase coexistence region for the 'standard' phase transition approach at generally higher density also the typical NJL-like shape of the phase boundaries is different compared to vBag. For vBag it generally increases with increasing temperature. This is related to the temperature dependence of $B_{\rm dc}$ which is shown in the upper panel of Fig.~\ref{fig:phase_vbag}. It decreases with increasing temperature. This behavior is determined by the hadronic EoS. We interpret $B_{\rm dc}$ as binding energy of quarks in confined matter. The associated additional pressure, in particular at low temperatures, shifts the phase boundaries towards lower densities compared to the 'standard' phase transition approach. Note that with increasing temperature both approaches coincide, i.e. $\mu_{\rm B, \rm dc}\rightarrow\mu_{\rm B,\chi}$, due to the decrease of $B_{\rm dc}$. For illustration we also plot curves of constant entropy per baryon in Fig.~\ref{fig:phase} (gray solid lines) for some selected values from $s=1-20$~k$_{\rm B}$/baryon. It results in a sudden decrease of $T$ at the phase boundary $\mu_{\rm B,\chi}^{rm vBag}$ respectively $\mu_{\rm B,\rm dc}^{\rm NJL}$. This feature is a result of the higher entropy per baryon in the quark phase and can be related to a decreasing phase transition pressure with increasing temperatures by the Clausius-Clapeyron equation. A phase transition with such a behavior is sometimes also called an entropic phase transition \cite{hempel13,iosilevskiy14,iosilevskiy15}. For simplicity here we did not calculate the behavior of $s$ inside the coexistence region (light blue bands in the bottom panels of Figs~\ref{fig:phase_vbag} and \ref{fig:phase_njl}). 

Towards high temperatures, in particular in excess of $T\simeq 100$~MeV, additional particles start to be excited at the hadronic side, e.g., pions as the lightest mesons. Equivalently thermal gluons are present in the quark phase. Currently none of these contributions are taken into account in vBag. Moreover, like all two-EoS approaches, vBag will always result in a 1st-order phase transition by construction. Hence we cannot make any statement regarding the existence of critical point(s). This would require a model where both quarks and hadrons are included as quasi-particle degrees of freedom in a unified description, or where hadrons can be formed as bound-states of quarks from their interactions. Only very few such models exist at present, and only from the former category \cite{Dexheimer:2010,Randrup:2010,Steinheimer:2011a,Steinheimer:2011b,Steinheimer:2014}.

Associated with a melting of the chiral condensate the chiral bag constant $B_\chi$ should have an explicit temperature dependence. Here we neglect such dependencies assuming the same value for $B_\chi$ for all temperatures. In Ref.~\cite{Klaehn:2016} we explored along this direction by varying $B_\chi$ and studying consequences for the phase diagram. With decreasing $B_\chi$ also $\mu_{\rm B,\chi}$ and $n_{\rm B, \chi}$ will reduce substantially. However, modeling the full medium dependence of $B_\chi$ requires a more elaborate approach than our phenomenological model, which we leave for future investigations. Moreover, towards high temperatures -- in particular in excess of $T\simeq 100$~MeV -- and low densities our hadronic EoS HS(DD2) is not physical anymore since thermally excited pions and hadronic resonance are not taken into account. 

\section{Summary}
\label{summary}

In this article we use the symmetric matter 2-flavor vBag approach to model quark matter. It takes  D$\chi$SB explicitly into account via the chiral bag constant $B_\chi$ and impose simultaneous chiral symmetry restoration and deconfienement. The latter condition connects  vBag to a given hadronic EoS in terms of the deconfinement bag constant ($B_{\rm dc}$). Furthermore, repulsive vector interactions are considered which stiffen the EoS towards high density.

A natural consequence of the extension to finite temperatures is the emergence of an implicit temperature dependence of the deconfinement bag constant, $B_\text{dc}(T)$, which is determined by the pressure of the hadronic EoS at the chiral transition. Corrections to entropy and energy density arise from a thermodynamically consistent treatment, while the phase diagram is independent from these corrections. As a consequence we find that simultaneous chiral symmetry restoration and deconfinement make the quark and hadronic phases more similar regarding their temperature dependences.

The future heavy-ion collider facilities at NICA in Dubna (Russia) and FAIR at the GSI in Darmstadt (Germany) will probe the state of matter at large chemical potentials. From these programs it may be possible to deduce whether simultaneous D$\chi$SB and (de)confinement is a feature of QCD at finite density. This is to be complemented by theory, e.g., simulations of heavy-ion collisions based on model EoS which are build on such assumption.

\bigskip

\setlength{\parindent}{0cm}

{\small
We  acknowledge support from the Polish National Science Center (NCN) under the grant numbers UMO-2011/02/A/ST2/ 00306 (TF) and UMO-2013/09/B/ST2/01560 (TK). MH acknowledges support from the Swiss National Science Foundation. Partial support comes from ``NewCompStar'', COST Action MP1304.
}

\bigskip

{\small
{\bf Open Access} This is an open access article distributed under the terms of the Creative Commons Attribution License (http://creativecommons.org/licenses/by/4.0), which permits unrestricted use, distribution, and reproduction in any medium, provided the original work is properly cited.
}


\end{document}